\documentclass[twocolumn,prb,superscriptaddress]{revtex4}
\usepackage{graphicx}
\usepackage{dcolumn}
\usepackage{bm}
\usepackage{amsmath}
\usepackage{times}
\usepackage{color}
\usepackage[breaklinks=true,colorlinks,citecolor=red,linkcolor=blue,urlcolor=red]{hyperref}

\makeatletter

\newcommand{\Rmnum}[1]{\expandafter\@slowromancap\romannumeral #1@}
\makeatother

\begin{document}

\title{Enhanced magnetic and optical properties of oxygen deficient TiO$_{2-\delta}$ nanoparticles synthesized by environment-friendly green route using whole plant extract of {\it{Phyllanthus niruri}}}

\author{Latika Mishra}
\email{itslatika@gmail.com}
\affiliation{Department of Zoology, Ranchi University, Ranchi 834008, India}

\author{Vinod Kumar Dwivedi$^\ast$}
\email{vinodd.iitbombay@gmail.com}
\affiliation{Department of Materials Science and Engineering, Tel Aviv University, Tel Aviv 6997801, Israel}

\author{Vishal Kumar Chakradhary}
\affiliation{RF Nanocomposites Pvt. Ltd., Startup Incubation and Innovation Centre (SIIC), Indian Institute of Technology Kanpur, Kanpur 208016, India}

\author{Akila G. Prabhudessai}
\affiliation{Department of Physics, Indian Institute of Science, Bengaluru 560012, India}

\author{Shamshun Nehar}
\affiliation{Department of Zoology, Ranchi University, Ranchi 834008, India}


\begin{abstract}

We report magnetic, optical and oxidation states of oxygen deficient TiO$_{2-\delta}$ nanoparticles (NPs) synthesized by environment-friendly green route using {\it{Phyllanthus niruri}} (PN) whole plant extract instead of leaf extract. Rietveld refinement of room temperature XRD pattern confirms the formation of pure phase anatase TiO$_2$ crystals in a tetragonal structure with space group I41/amd. TEM and SEM microstructure shows agglomerated spherical shape NPs exhibiting average particle size $\sim$ 35~nm. FTIR result confirms the presence of biomolecules and functional group attached to the surface of TiO$_2$ NPs. The core level XPS of O-1s and Ti-2p confirms the presence of oxygen vacancies that leads to the mixed oxidation states of Ti (Ti$^{4+}$ and Ti$^{3+}$). UV-vis result shows a strong absorption peak ($\sim$ 250~nm) along with reduced optical band gap energy E$_g$ $\sim$ 2.75~eV, possibly arises due to the surface plasmon resonance (SPR) caused by {\it{lower band gap energy}} emerging from oxygen vacancies. Magnetization as a function of applied magnetic field shows ferromagnetic nature at room temperature [M$_S$ $\sim$ 0.029~emu/g and H$_C$ $\sim$ 0.0143~T]. The observed ferromagnetic behaviour can be understood by {\it{virtual hopping}} of electrons from Ti$^{3+}$(3d$^1$) to Ti$^{4+}$(3d$^0$)-sites, however, vice versa is prohibited.

\end{abstract}

\maketitle
\section{Introduction}
The recent advancement at the interface of biological systems and materials science has gained attention worldwide to explore the new nano-structured bio-materials. It show exotic properties that are different from their bulk counterpart~\cite{Rana,Chavali,Negrescu,Vinod21,Vinod22,Vinod23}. Among all metal oxides, TiO$_2$ in its nanoparticles (NPs) shape has particular interest because of its photocatalytic properties, non-toxicity and chemical stability. The use of TiO$_2$ NPs have been expected for various uses in the field of sensor, solar cell, photocatalysis, antimicrobial, antibiotic, cosmetic industry, electrochemical devices, spintronics, antibacterial and so on~\cite{Aslam,Singh,Kumar,Ullah,Bhullar,Narayanan,Kashale}. It is evident that TiO$_2$ NPs have been utilized in cancer photothermal therapy~\cite{Sargazi} and proposed for the use of non-radiative recombination. Moreover, it has been reported as potential candidate for wastewater treatment as it has capability to degrade and detoxify the toxic waste from the waste water~\cite{Lazar}.

Conventionally, TiO$_2$ NPs are synthesized by chemical route~\cite{Parras,Qian,Dhakshinamoorthy1,Dhakshinamoorthy2}. The harmful chemicals are used in this method. Therefore, the exposure of harmful chemicals to the test organisms and environment could lead to the distinct level of toxicity and restrict their applications~\cite{Shi}. Thus, to overcome the toxicity of NPs, green synthesis route has been explored. Green route does not require hazardous chemicals, expensive equipment, and high temperatures etc. In this method, different plant extracts and microorganisms have been used to synthesize the TiO$_2$ NPs~\cite{Mao,Sunny,Rajaram,Azad}. The use of plant photochemicals provide a catalyst for hydrolysis reaction, stabilizing capping agent and a cover to stop agglomeration during synthesis.

{\it{Phyllanthus niruri}} (PN) is a member of Euphorbiaceae family. It is an important herb in Indian Ayurvedic medicine over thousands of years. PN has been utilized for the treatment of various stress-related diseases and disorders such as diabetes, jaundice, asthma, diarrhea, hepatic disease, kidney aliments, improve eyesight, constipation, skin disorder, ulcers, lung-related disease, malaria, respiratory disorder, ringworm, etc. It is also used in fever, and hypertension~\cite{Lee1,Bagalkotkar}. In addition, the numerous NPs made by green synthesis route using PN leaf extract have shown a potential for their use as antifungal, antidiabetic and antibacterial agents~\cite{Khanna}. The extract of PN plant serves as a reducing, stabilizing and capping agent and that leads to the depletion of Ti ions to the formation of TiO$_2$ NPs.

Recently, CuO nanoparticles has been prepared by same group~\cite{Mishra} using whole plant extract of PN. It is interesting to mention that, these CuO NPs shows a core-shall like magnetic structure and optical band gap is enhanced compared to their bulk counterparts. S. Shanavas, et al.,~\cite{Shanavas} have prepared TiO$_2$ NPs (average particles size $\sim$ 32~nm) by green route using PN leaf extract. The estimated optical band gap energy (E$_g$) was found to be 3.16~eV. A. Panneerselvam, et al.,~\cite{Panneerselvam1} have studied the removal of methyl orange dye using various parameters in TiO$_2$ NPs ($\sim$ 20~nm) synthesized using PN leaf extract. A. K. Shimi, et al.,~\cite{Shimi1} have investigated the catalytic activity of TiO$_2$ NPs (particle size $\sim$ 23~nm, E$_g$ $\sim$ 3.16~eV) prepared by green route using leaf extract of PN. N. Jayan, et al.,~\cite{Jayan} have shown the adsorption of heavy metal on TiO$_2$ NPs (various sizes of nanoparticles $\sim$ 20-40~nm) surface synthesized using {\it{Phyllanthus acidus}} extract. Despite these advances~\cite{Shanavas,Panneerselvam1,Shimi1,Jayan}, there are {\it{no reports}} on magnetic properties, experimental evidence of oxidation states and reduction of optical band gap energy E$_g$ along with strong absorption peak of TiO$_{2-\delta}$ NPs synthesized by green route using PN whole plant extract.

Here, we present structural, optical and magnetic properties of oxygen deficient TiO$_{2-\delta}$ NPs synthesized by green route using whole plant extract of PN over leaf extract. The reduction of particle size decreases the oxidation states of Ti from Ti$^{4+}$ to Ti$^{3+}$. As a result, TiO$_{2-\delta}$ NPs exhibit:- (I) significant reduction in optical band gap energy (E$_g$ $\sim$ 2.75~eV) with strong absorption peak ($\sim$ 250~nm) in UV-vis spectrum compared to bulk polycrystalline TiO$_2$ (E$_g$ $\sim$ 3.2~eV) and PN leaf extract, and (II) room temperature ferromagnetism with two order magnitude enhancement of saturation magnetization (M$_S$ $\sim$ 0.029~emu/g, H$_C$ $\sim$ 0.0143~T) compared to TiO$_2$ NPs synthesized by other plants extract. These properties are {\it{entirely lacking}} in literature of TiO$_2$ NPs synthesized by green route using PN leaf extract~\cite{Shanavas,Panneerselvam1,Shimi1,Jayan}. As Compared to leaf extract of PN, the extracts of {\bf{whole plant}} exhibit distinct metabolites like phyllanthin and hypphyllanthin, corilagin, ellagic acid, geraniin, gallic acid. These phytochemicals are attached to the surface of NPs, may lead to the creation of oxygen vacancies. As a result, it gives rise to mixed oxidation states of Ti that leads to enhanced optical and magnetic properties of current TiO$_{2-\delta}$ NPs.    

\section{Experimental Method}

\subsection{Materials}
The preservative free and 100\% natural powder of whole plant PN were purchased from Sierra India Organics. Tetrabutyl titanate (TT) TiC$_{16}$H$_{36}$O$_4$ [purum, $\geq$ 97.0\% (gravimetric), liquid form], and ethanol (CH$_3$CH$_2$OH) [$\geq$ 99.5\% liquid form] were purchased from Sigma Aldrich.

\subsection{Preparation of PN whole plant extract}
The whole plant PN powder was dried at room temperature in the shade. Whole plant PN dried powder (5~g) were immersed in 200~mL deionized (DI) water and mixed using magnetic stirrer for 10~hrs at 60~$^0$C. Further, the solution was cool down to room temperature. The acquired PN whole plant extract was obtained by filtering the cooled solution using Whatman filter paper. The resultant PN whole plant extract was placed in refrigerator for the preparation of TiO$_2$ NPs.

\subsection{Preparation of TiO$_2$ NPs by green synthesis route} 
TiO$_2$ NPs were synthesized by green route using PN whole plant extract following the protocol reported elsewhere~\cite{Mishra}. The TT solution (40~mL) was immersed in 150~mL ethanol at room temperature. The solution was mixed using magnetic stirrer for 10~hrs at 60~$^0$C. The resultant TT solution was mixed to whole plant extract followed by continuous stirring for 12~hrs at temperature 70~$^0$C. The hydrolysis of TT is the central mechanism behind the formation of TiO$_2$ NPs. The whole plant extract present in solution, act as a stabilizing agent helps to prevent agglomeration and achieving the desired shape and size of the TiO$_2$ NPs. The colour of resulting solution (formed after mixing TT and whole plant extract of PN) changes from brown to whitish brown and finally yellowish. It indicates the formation of TiO$_2$ NPs. Further, solution was dried at 120~$^0$C for 8~hrs. The obtained chunk was crushed and ground using mortar and pestle to make fine powder. The powder was kept in alumina crucible and calcined at 570~$^0$C for 2~hr in muffle furnace. The colour of powder changes from yellow to white, confirms the probable formation of TiO$_2$ NPs.

\subsection{Characterization} 
To determine the crystal structure and phase formation of TiO$_2$ NPs, powder x-ray diffraction (XRD) pattern was measured at room temperature using PANalytical XPertPRO diffractometer equipped with Cu-K$_\alpha$ radiation ($\lambda$ = 1.54056~\AA). The XRD pattern is collected in 2$\theta$ range from 20 to 80$^0$ with a step size 0.02$^0$. The crystallinity, particle size and distribution of TiO$_2$ NPs were verified using high resolution transmission electron microscopy (HR-TEM) model FEI Titan G2 60-300, performed at beam voltage 300~kV. The sample for TEM measurement were prepared by dispersing the TiO$_2$ NPs in Isopropyl alcohol following the ultrasonication to decrease the agglomeration. The resultant solution was drop-cast onto a copper mesh. The energy dispersive x-ray spectrometry (EDX) and particle size distribution was also performed using field emission scanning electron microscope (FE-SEM) Model - JSM-7100F; JEOL. The presence of different functional groups attached to the surface of NPs was determined using Fourier Transform Infrared (FTIR) spectroscopy (model: Frontier FIR/MIR, from Perkin Elmer, USA) measured in the wavenumber range of 4000-500~cm$^{-1}$. It was averaged over 32 scans with a special resolution of 2~cm$^{-1}$. Attenuated Total Reflectance (ATR) technique was used to perform measurement on pellet of TiO$_2$ NPs mixed with KBr. Raman spectra is measured using Raman spectrometer (Model: Princeton Instruments Acton Spectra Pro 2500i) equipped with laser excitation wavelength 532~nm. Spectrum were measured at three distinct positions. X-ray photoelectron spectroscopy (XPS) model PHI 5000 Versa Prob II, FEI Inc was used to record the XPS spectrum with step size 0.05~eV and an energy resolution of 0.02~eV. The optical absorbance spectra were measured using Ultra violet visible (UV-vis) spectrophotometer (model: Eppendorf BioSpectrometers in wavelength range of 100 to 800~nm. The magnetic properties of TiO$_2$ NPs were carried out using a vibrating sample magnetometer (Princeton VSM model-150) at room temperature in the magnetic field of range $\pm$ 1~Tesla.

 \section{Results and Discussion}

\begin{figure}
	\centering
	\includegraphics[width=\linewidth]{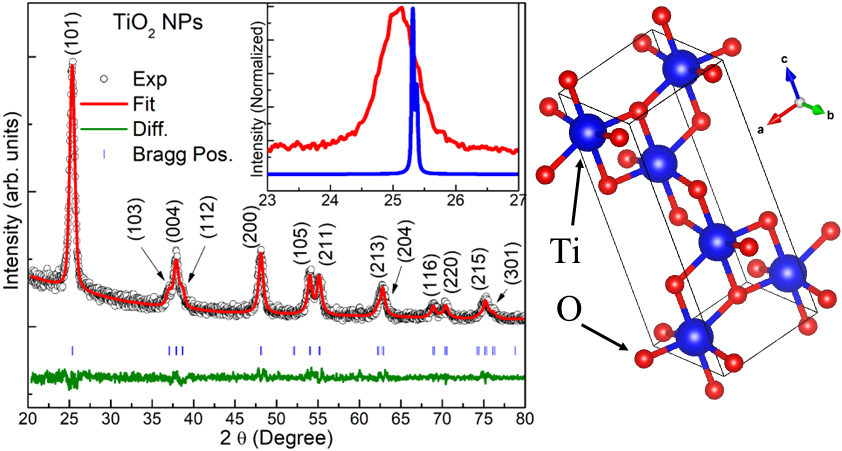}\\
	\caption{Left panel - Rietveld refined XRD pattern of TiO$_2$ NPs; inset shows broadness of most intense peaks of TiO$_2$ NPs (red line) and reference sample polycrystalline TiO$_2$ (blue line) at normalized intensity scale. Right panel - crystal structure of anatase TiO$_2$ NPs obtained by VESTA software. The smaller atom oxygen O and larger atom titanium Ti is shown by red and blue solid sphere, respectively.}\label{fig:xrd}
\end{figure}

\begin{table}
	\caption{The refined parameters of XRD pattern. \label{XRDrefined}}
	\begin{center}
		\begin{tabular}{c c c c c }
			\hline
			Atoms&x&y&z&Occupancy\\
			\hline
			Ti&0.0000&0.7500&0.1250&0.71\\
			\hline
			O&0.0000&0.7500&0.3335&0.57\\
			\hline
		\end{tabular}
	\end{center}
\end{table}

Figure~\ref{fig:xrd} (left panel) reveals XRD pattern of green synthesized TiO$_2$ NPs using extract of whole plant PN.  
FULLPROF software using Rietveld refinement method were employed to analyze the XRD pattern. The refined parameter goodness of fit is defined as $\chi^2$ = $[\frac{R_{wp}}{R_{exp}}]^2$, where R$_{wp}$ and R$_{exp}$ is expected weighed profile factor and observed experimental weighed profile factor, respectively. Refinement result shows that all observed peaks can be indexed to pure anatase phase tetragonal structure with space group I41/amd. The obtained refined parameters are $\chi^2$ = 1.37, R$_{wp}$ = 12.99 and R$_{exp}$ = 11.1, a = b = 3.7819~\AA, c = 9.4979~\AA and $\alpha$ = $\beta$ = $\gamma$ = 90$^0$. The small value of goodness fit suggests the single phase formation of highly crystalline TiO$_2$ NPs. It occurs likely due to high reduction potential of PN whole plant extract that prevents the formation of any amorphous phase. It can also be seen that refined crystal structure demonstrates occupancy less than one for both Ti and O atoms [Table~\ref{XRDrefined}], suggests the probable consequent development of oxygen vacancies~\cite{Sarkar}. Inset of left panel Fig~\ref{fig:xrd} shows the most intense peak of TiO$_2$ NPs (red line) and polycrystalline reference sample TiO$_2$ (blue line). It is obvious that the broadness (full width at half maximum) of TiO$_2$ NPs is much larger than reference bulk polycrystalline sample, confirming the formation of NPs. The average crystallite size/diameter of particle is calculated using Debye-Scherrer's formula D = $\frac{(0.94\lambda)}{(\beta cos\theta)}$, where $\lambda$, $\beta$, and $\theta$ represents the wavelength of X-ray, full width at half maximum in radians of XRD peaks, and diffraction angle in radians, respectively. The estimated average crystallite size turns out to be 35~nm. The right panel of Fig.~\ref{fig:xrd} reveals the crystal structure of tetragonal anatase TiO$_2$ NPs. 

\begin{figure}
	\centering
	\includegraphics[width=\linewidth]{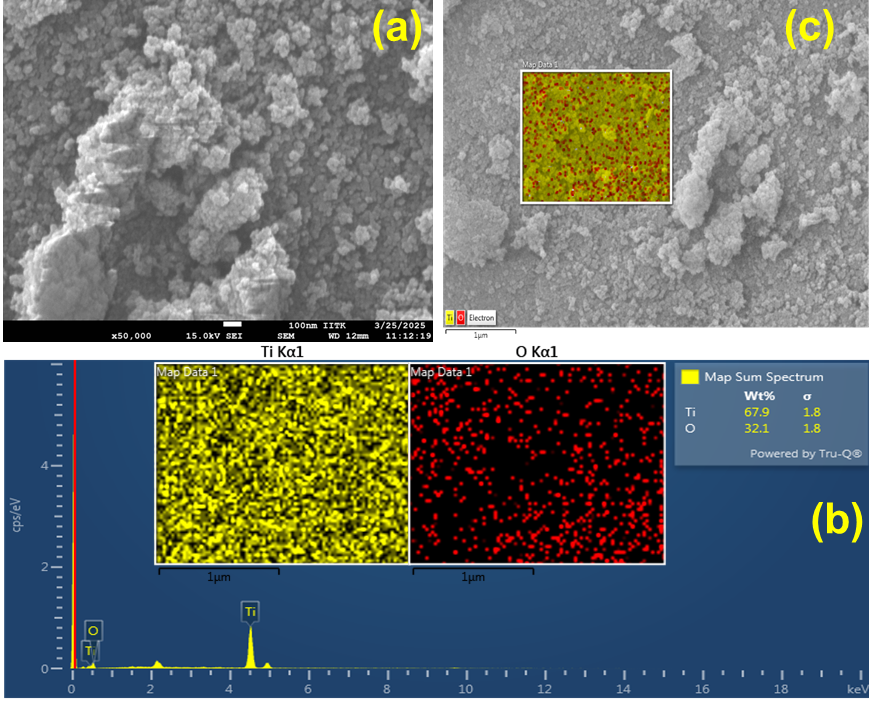}\\
	\caption{(a) SEM micrograph of green synthesized TiO$_{2-\delta}$ NPs, (b) Energy dispersive X-ray (EDX) spectra of TiO$_{2-\delta}$ NPs; inset shows titanium (golden color) and oxygen (red color) elemental color mapping and (c) Selected area at which EDX elemental color mapping were measured.}\label{fig:SEM}
\end{figure}

\begin{figure}
	\centering
	\includegraphics[width=\linewidth]{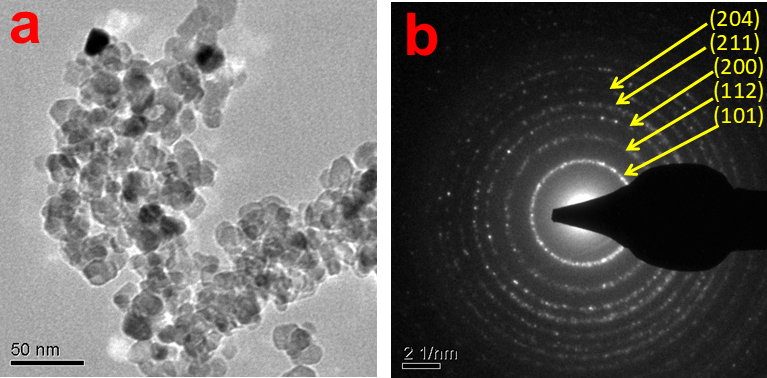}\\
	\caption{(a) Transmission electron microscopy (TEM) image, and (b) selected area electron diffraction pattern of TiO$_{2-\delta}$ NPs synthesized by green route.}\label{fig:TEM}
\end{figure}

The size and surface morphology of TiO$_2$ NPs were scanned at magnification x~50000 and applied voltage 15~kV by scanning electron microscope (SEM) shown in Fig.~\ref{fig:SEM}a. SEM image reveals even distribution of spherical NPs across the surface with average particles size of 35~nm. It shows rough, agglomerated clusters and irregular porous surface of distributed NPs. The estimated average particle size of NPs using SEM show a close agreement with average size calculated using XRD. The smaller size spherical nanoparticles exhibit larger surface to volume ratio provides a platform for plant biomolecules to form a layer. These layers of biomolecules attract other nanoparticles, leading to the agglomeration of nanoparticles. Figure~\ref{fig:SEM}b displays the energy dispersive X-ray (EDX) spectra of green synthesized TiO$_2$ NPs measured across the large area [Fig.~\ref{fig:SEM}c]. It confirms the presence of expected elements Ti and O across the surface. Strong absorption peaks are related to Ti and O due to TiO$_2$ NPs. The small peaks less in intensity  centered at $\sim$ 0.65, 1.3, 1.5, and 1.75~keV are possibly associated to plant molecules consistent with the litterateurs~\cite{Panneerselvam1,Jayan}. It is obvious that the experimental elemental mass ratio [$\frac{Ti}{O}$ $\sim$ 2.1] shown by EDX is much larger than the theoretical value (1.5), suggesting oxygen deficiency in TiO$_2$ NPs, i.e., TiO$_{2-\delta}$. Transmission electron microscopy (TEM) micrograph [Figure~\ref{fig:TEM}a] reveals spherical shape NPs with average particle size $\sim$ 35~nm. It is in fair agreement with estimated particle size from XRD pattern using Debye-Scherrer's formula and SEM data. Figure~\ref{fig:TEM}b shows the selected area electron diffraction pattern of TiO$_{2-\delta}$ NPs. It shows clear separated rings made of bright small spots. Rings can be indexed from the d-spacing value of crystal planes determined using software Image J. The estimated planes are in good agreement with the planes found from refined XRD pattern [Fig.~\ref{fig:xrd}a]. Therefore, the obtained rings in SAED pattern are attributed to various crystal planes of tetragonal anatase phase, suggesting the well crystalline nature of TiO$_{2-\delta}$ NPs.

\begin{figure}
	\centering
	\includegraphics[width=\linewidth]{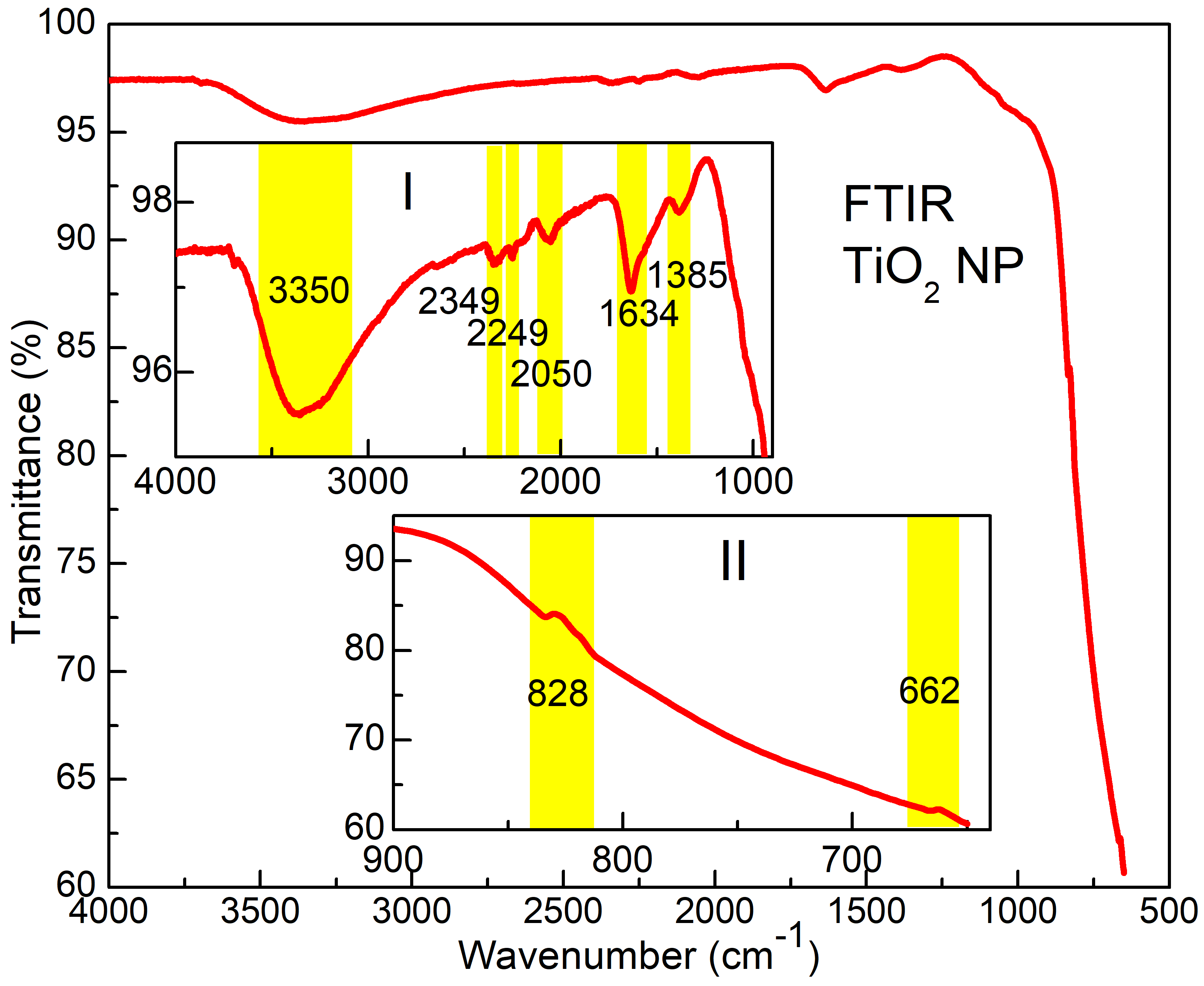}\\
	\caption{Fourier-transform infrared spectroscopy (FTIR) spectrum of green synthesized TiO$_{2-\delta}$ NPs. Inset I and II show enlarged image of the same.}\label{fig:FTIR}
\end{figure}

The reduction of TiO$_{2-\delta}$ NPs due to the presence of various plant active functional groups reside on the surface of nanoparticles were investigated using Fourier-transform infrared spectroscopy (FTIR) technique. Figure~\ref{fig:FTIR} shows FTIR spectrum of green synthesized TiO$_{2-\delta}$ NPs. Spectrum displays distinct bands in different regions [inset I and II of Fig.~\ref{fig:FTIR}]. The FTIR transmittance prominent peak centered at $\sim$ 3350~cm$^{-1}$ [inset I of Fig~\ref{fig:FTIR}] belongs to the O-H stretching vibrations in phenol and alcohol compounds. The peak located at $\sim$ 2349~cm$^{-1}$ represents C-H stretching vibrations of the aromatics group favours the formation of organic precipitation during the time of synthesis process. A less prominent peak centered at $\sim$ 2249~cm$^{-1}$ can be seen. It is attributed to Si-H stretching vibration related to organosilicon compounds. The peaks situated at $\sim$ 2050~cm$^{-1}$ corresponds to N=C=S stretching vibrations in isothiocynate group. The prominent peak at $\sim$ 1634~cm$^{-1}$ associated to C-H or C=C stretching vibrations of alkenes. A small peak can also be seen at $\sim$ 1385~cm$^{-1}$ related to the stretching vibration of C-F in alkyl halides. Therefore, the identified various functional groups in the PN whole plant extract likely act as capping agents during the synthesis process to stabilize the tetragonal anatase phase of TiO$_2$. Fig.~\ref{fig:FTIR} inset II shows the FTIR spectrum of TiO$_{2-\delta}$ NPs in the low wave number region (900-500~cm$^{-1}$). The low wavenumber region is a significant area to identify the vibration of metal-oxide bonds especially in the range 900-500~cm$^{-1}$. The peak centered at wavenumber $\sim$ 828~cm$^{-1}$ and $\sim$ 662~cm$^{-1}$ belongs to the stretching vibrations of Ti-O-Ti and Ti-O bonds in TiO$_{2-\delta}$ NPs. The above observed functional groups suggest the existence of plant biomolecules on the surface of nanoparticles. It promotes and stabilize the formation of Ti-O bond during synthesis.

\begin{figure}
	\centering
	\includegraphics[width=\linewidth]{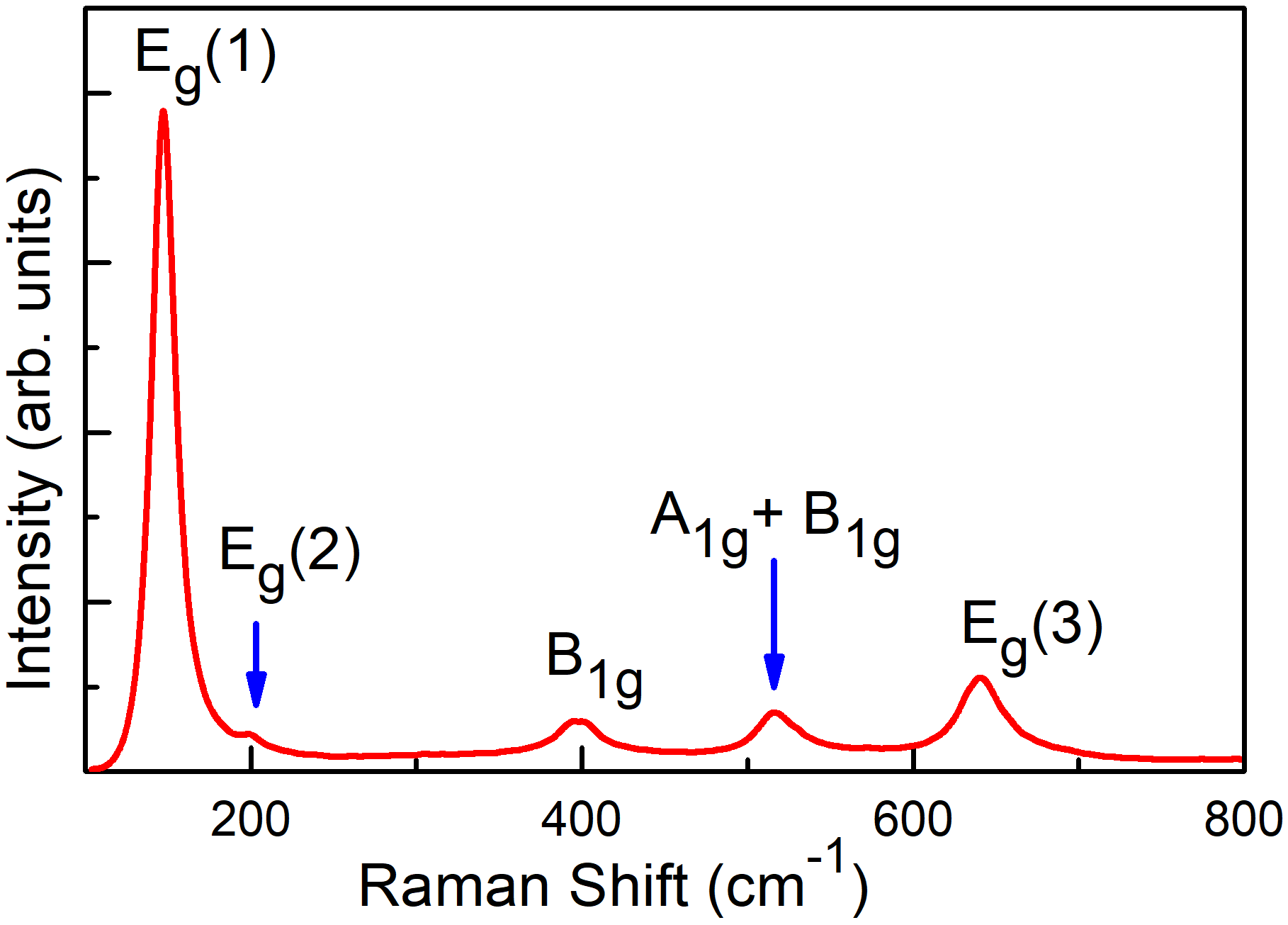}\\
	\caption{Raman spectrum of green synthesized TiO$_{2-\delta}$ NPs measured in the range 104-10000~cm$^{-1}$ using excitation wavelength of laser $\lambda$ = 532~nm.}\label{fig:Raman}
\end{figure}

To determine the structure, defects and nanocrystalline phase, Raman scattering measurement was carried out at room temperature using the laser excitation wavelength $\lambda$ = 532~nm, laser power = 40~mW and grating = 600. Figure~\ref{fig:Raman} shows Raman spectrum of TiO$_2$ NPs measured in wavenumber ranging 100 - 1000~cm$^{-1}$. It is obvious that Raman spectrum exhibits five peaks. The Raman spectrum of anatase phase TiO$_2$ contains five Raman active modes in the vibrational spectrum; three E$_g$ modes and two B$_{1g}$ modes. The peaks centered at wavenumber $\sim$ 147, 199 and 641~cm$^{-1}$ are attributed to the modes E$_g$(1), E$_g$(2) and E$_g$(3), respectively, while the peaks situated at wavenumber $\sim$ 398 and 517~cm$^{-1}$ are related to B$_{1g}$ and A$_{1g}$ + B$_{1g}$ modes, respectively. The observed peaks are consistent with literature~\cite{Mao,Swamy,Parker2,Kumari}. It confirms the formation of anatase TiO$_2$. The Raman mode E$_g$(1) emerges due to the vibrational modes of oxygen anions (O$^{2-}$) and Ti-O bonds. The FWHM of peaks are relatively larger and shifted towards higher wavenumber side compared to bulk TiO$_2$. It suggests the presence of oxygen vacancies/defects (Ti$^{3+}$) in TiO$_2$ NPs. The enhancement in broadening and shifting in Raman peaks have been reported when one approaches from the bulk polycrystalline to nanoparticles phase of TiO$_2$~\cite{Swamy,Parker2}. We will come back to this point (oxygen vacancies/defects (Ti$^{3+}$)) shortly hereafter.       

\begin{figure}
	\centering
	\includegraphics[width=\linewidth]{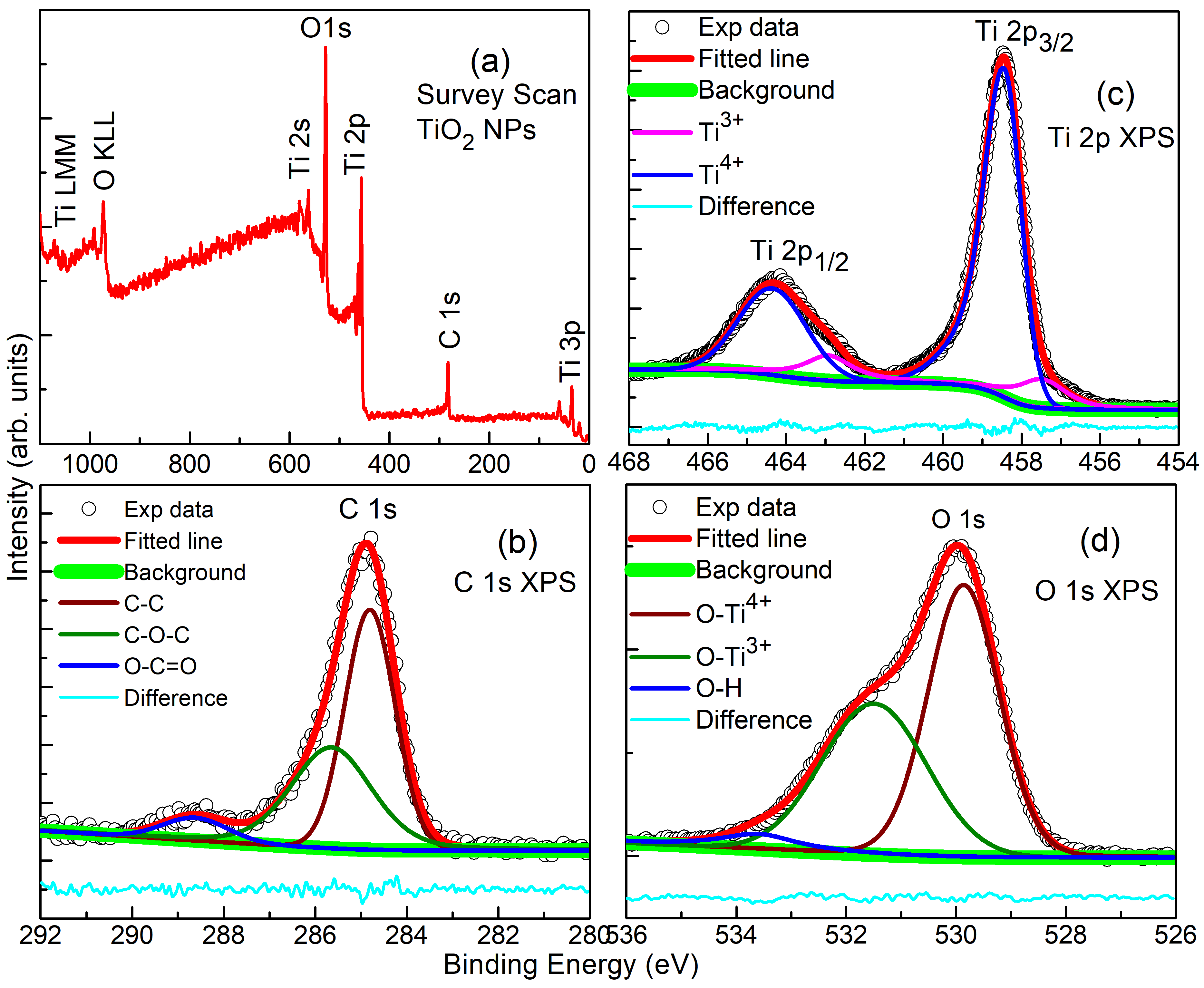}\\
	\caption{(a) XPS survey spectra of TiO$_{2-\delta}$ NPs. Deconvoluted XPS spectra of (b) C 1s, (c) Ti 2p, and (d) O 1s for TiO$_{2-\delta}$ NPs prepared by green route using PN whole plant extract.}\label{fig:XPS}
\end{figure}

X-ray photoelectron spectroscopy (XPS) is a highly sensitive technique to explore the surface research to instrument limit ($\sim$ 10~nm). Therefore, XPS measurement was carried to investigate the chemical environment and surface states of green synthesized TiO$_2$ NPs. Figure~\ref{fig:XPS}a shows survey scan spectra of TiO$_2$ NPs, suggesting the presence of desired element content on the surface. Figure~\ref{fig:XPS}b reveals C-1s XPS spectrum. Spectrum can be fitted with three constituent peaks. The peak centered at 284.81~eV is attributed to the C-C neutral bonds. The peaks located at 285.83~eV and 288.62~eV are related to the C-O-C and O-C=O bond, respectively of the carbon species. Generally, the samples exposed to the open atmosphere acquire a detectable amount of carbon contamination adventitiously up to few nanometer depth ($\sim$ 1-2~nm). The C-1s XPS peak situated at 284.81~eV can be used as a reference peak for other XPS peaks. 

\begin{table}
	\caption{Ti-2p XPS; percentage of total area is represented by \%. \label{Tixps}}
	\begin{center}
		\begin{tabular}{c c c c c }
			\hline
			&Peak 1&Peak 2&Peak 3&Peak 4\\
			\hline
			Position (eV)&457.2 &458.58&462.96&464.31\\
			\hline
			\% &13&51&9&27\\
			\hline
		\end{tabular}
	\end{center}
\end{table}

In order to investigate the oxidation states of Ti, we measured the XPS core level spectrum. Figure~\ref{fig:XPS}c shows Ti-2p core level XPS spectrum of TiO$_2$ NPs. XPS peaks can be indexed following the protocol reported elsewhere~\cite{Sanjines,Thejas,Kumari1}. We have fitted Ti-2p core level XPS spectrum using the asymmetric Lorentz-Gauss sum function with Shirley background using software XPS peakfit4.1. For a good fit of Ti-2p core level XPS, experimental data indicates that two oxidation states of Ti are needed. For Ti$^{4+}$, 2p$_{3/2}$ and 2p$_{1/2}$ electronic states arise at binding energies $\sim$ 458.58~eV (peak 2) and 464.31~eV (peak 4) with significant spin-orbit splitting components ($\bigtriangleup$ = 5.73~eV), respectively, which are shown by the blue solid in Fig.~\ref{fig:XPS}c. Similarly, a lower oxidation state Ti$^{3+}$ is identified at binding energies $\sim$ 457.20~eV (peak 1) and 462.96~eV (peak 3) displayed by solid pink line [Fig.~\ref{fig:XPS}c] for the electronic states 2p$_{3/2}$ and 2p$_{1/2}$ with $\bigtriangleup$ = 5.76~eV, respectively. The  fitting parameters are shown in Table~\ref{Tixps}. It suggest the coexistence of mixed oxidation states of Ti i.e., Ti$^{3+}$ (small contribution) and Ti$^{4+}$ (major contribution) for the TiO$_2$ NPs synthesized by green route using PN whole plant extract. Although, Ti$^{3+}$ cation is less prominent in Ti-2p XPS (Fig.~\ref{fig:XPS}c), the percentage of total peak area for Ti$^{3+}$ (peak 1) exhibit significant value (13\%). The presence of Ti$^{3+}$ in TiO$_2$ NPs might be attributed to the nonstoichiometry such as deficiency of transition metal elements as suggested by EDX and Raman scattering results.

\begin{table}
	\caption{O-1s XPS; percentage of total area is represented by \%. \label{Oxps}}
	\begin{center}
		\begin{tabular}{c c c c }
			\hline
			&Peak 1&Peak 2&Peak 3\\
			\hline
			Position (eV)&529.86 &531.5&533.5\\
			\hline
			\% &48&49&3\\
			\hline
		\end{tabular}
	\end{center}
\end{table}

It is known that multiple valance states affect the local surroundings of metal-oxygen bonds. Therefore, further support for the mixed oxidation states of Ti can be seen in the O-1s core level XPS spectrum [Fig.~\ref{fig:XPS}d]. The O-1s spectrum is asymmetric in shape and fitted with three peaks related to the oxide component. Fitting parameters are shown in Table~\ref{Oxps}. The peak situated at binding energy $\sim$ 529.86~eV is attributed to an oxygen lattice of TiO$_2$ NPs while peak centered at $\sim$ 531.49~eV is likely due to oxygen vacancies or defects such as oxygen in Ti$_2$O$_3$ (O-Ti$^{3+}$/Ti$^{4+}$). The percentage of total peak area for O-Ti$^{3+}$ (peak2) was found to be 49\%, favours the strong presence of oxygen vacancies/Ti$^{3+}$. The peak located at $\sim$ 533.2~eV corresponds to organic C-O of green synthesized TiO$_2$ NPs using PN whole plant extract.

\begin{figure}
	\centering
	\includegraphics[width=\linewidth]{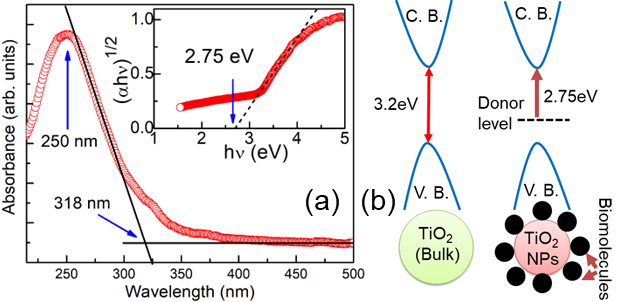}\\
	\caption{(a) UV-vis absorption spectrum of green synthesized TiO$_{2-\delta}$ NPs. Inset shows Tauc plot for the same derived from reflectance spectrum. (b) Schematic demonstration of charge transfer for TiO$_{2-\delta}$ NPs from donor level (dashed line) to conduction band (C.B.) giving rise to decreased E$_g$; V.B. stand for valance band.}\label{fig:UVvis}
\end{figure}

Further, to study the optical properties and electronic states of TiO$_{2-\delta}$ NPs synthesized by green route using PN whole plant extract, we measured absorption spectrum using Ultraviolet-visible (UV-vis) spectroscopy shown in Fig.~\ref{fig:UVvis}a. Absorption band edge starts at wavelength $\sim$ 318~nm shows a steep enhancement in absorption possibly due to band to band transition. A strong absorption peak can be seen at wavelength $\sim$ 250~nm. It is {\it{completely absent}} in the previously reported literatures of TiO$_2$ NPs synthesized by green route using PN leaf extract~\cite{Shanavas,Panneerselvam1,Shimi1,Jayan}, possibly associated to the formation of surface plasmon resonance (SPR) emerging from oxygen vacancies. Similar phenomenon have been reported in oxygen deficient  TiO$_2$ NPs synthesized by various distinct methods~\cite{Li2,Li3,Yang1,Yang2,Pan,Song}. Surface plasmons exhibit lower energy as compared to its bulk counterpart plasmons. In TiO$_{2-\delta}$ NPs, plant active bio-molecules attached to the surface of NPs [Fig.~\ref{fig:UVvis}b] reconstructs the electronic structure. Both valence band (VB) and conduction band (CB) come closer to each other, as a result charge transfer can takes place betwenn VB and CB. When light falls on the nanoparticles surface at a particular frequency, the momentums of surface plasmons be the same as momentums of incident photons. In this condition, incident photons can transfer all energy to the free electrons resides at the interface known as resonant excitation of surface plasmons. It controls the maximum in absorption spectra at specific frequency caused by the optical energy absorption.

The value of E$_g$ is estimated using Tauc equation $\alpha$h$\nu$ = A (h$\nu$ - E$_g$)$^n$, where $\alpha$, h$\nu$, and A are absorption coefficient, photon energy, and proportionality constant, respectively. Here exponent n is a constant depends on the type of transitions. For the values of {\it{n}} = 3, 3/2, 2, and 1/2, transition corresponds to the forbidden indirect, forbidden direct, allowed indirect, and allowed direct, respectively. Inset of Fig.~\ref{fig:UVvis}a shows Tauc's plot. The estimated value of E$_g$ found to be 2.75~eV with allowed direct transition (n = 1/2). Table~\ref{Parametercomparison} shows the comparison between the E$_g$ value for current TiO$_{2-\delta}$ NPs (synthesized using PN whole plant extract) and reported TiO$_2$ NPs (prepared using PN leaf extract and other plants). The estimated value of E$_g$ is smaller than the bandgap of bulk polycrystalline TiO$_2$ (3.2~eV)~\cite{Scanlon,Dey}. It can also be seen that E$_g$ value of TiO$_2$ NPs prepared using PN leaf extract and other plants extract exhibit higher value than current TiO$_{2-\delta}$ NPs. The reduction in optical band gap in current sample is possibly due to the presence of oxygen vacancies, generating pair of electrons~\cite{Dette}. One of the electron may reach to the vicinity of Ti$^{4+}$-site creating Ti$^{3+}$ centers. It leads to the formation of donor levels in the TiO$_{2-\delta}$ electronic band structure [Fig.~\ref{fig:UVvis}b]. Therefore, the transition of electrons between the new level (due to the presence of Ti$^{3+}$ along with Ti$^{4+}$) leads to the reduction of band gap. The band gap of Ti$_2$O$_3$ is 0.14~eV~\cite{Yu,Vijeta}. Therefore, presence of Ti$^{3+}$-cation would reduce the band gap in current sample. Quantitatively, Ti-2p XPS suggest 13\% contribution of Ti$^{3+}$-cation in current NPs, that reduces the band gap by the factor 0.42~eV [3.2x(13/100)]. As a result, the band gap is 3.2-0.42 = 2.78~eV consistent with current band gap 2.75~eV.

\begin{table*}
	\caption{Comparison of various parameters between current TiO$_{2-\delta}$ NPs synthesized by green route using whole plant extract and TiO$_2$ NPs synthesized by green route using other plants extract and conventional chemical route.\label{Parametercomparison}}
	\begin{center}
		\begin{tabular}{c c c c c c c c }
			\hline
			Extract&Method&Size&Absorption peak&E$_g$&M$_S$&References \\
			&&(nm)&(nm)&(nm)& (emu/g)& \\
			\hline
			{\it{Phyllanthus niruri}}&Green route&35&250, strong&{\bf{2.75}}&{\bf{0.029}}&{\bf{Current}}\\
			(PN) whole plant&&&&&&{\bf{work}} \\
			\hline
			PN leaf&Green route&32&absent&3.16&-&~\cite{Shanavas}\\
			\hline
			PN leaf &Green route&20.90&-&-&-&~\cite{Panneerselvam1}\\
			\hline
			PN leaf &Green route&23&absent&3.16&-&~\cite{Shimi1}\\
			\hline
			{\it{Phyllanthus}} &Green route&23&absent&3.16&-&~\cite{Jayan}\\
			acidus leaf&&&&&& \\
			\hline
			Citrus limon&Green route&15&-&-&-&~\cite{Singh}\\
			\hline
			Different&Green route&10-20&absent&3.0&-&~\cite{Kumar}\\
			tea leaf&&&&&& \\
			\hline
			Artocarpus &Green route&15-20&-&-&0.0002&~\cite{Ullah}\\
			heterophyllus&&&&&&\\
			\hline
			black pepper,&Green route&5.72,&-&-&0.004,&~\cite{Bhullar}\\
			coriander,&&5.57,&&&0.008& \\
			clove&&6.59&&&0.004&\\
			\hline
			Isopropyl alcohol&Chemical route&10&-&-&0.012&~\cite{Parras}\\
			\hline
			citric acid,&Chemical route&10-20&-&-&0.0086&~\cite{Qian} \\
			ethylene glycol&&&&&& \\
			\hline
			2-propanol,&Chemical route&13&-&-&0.00018&~\cite{Dhakshinamoorthy1}  \\
			acetic acid&&&&&&\\
			\hline
			2-propanol,&Chemical route&13&-&-&0.0006&~\cite{Dhakshinamoorthy2}  \\
			acetic acid&&&&&&\\
			\hline
			Ethanol&Chemical route&13.88&340&3.1&0.005&~\cite{Thejas}  \\
			\hline
		\end{tabular}
	\end{center}
\end{table*}

\begin{figure}
	\centering
	\includegraphics[width=\linewidth]{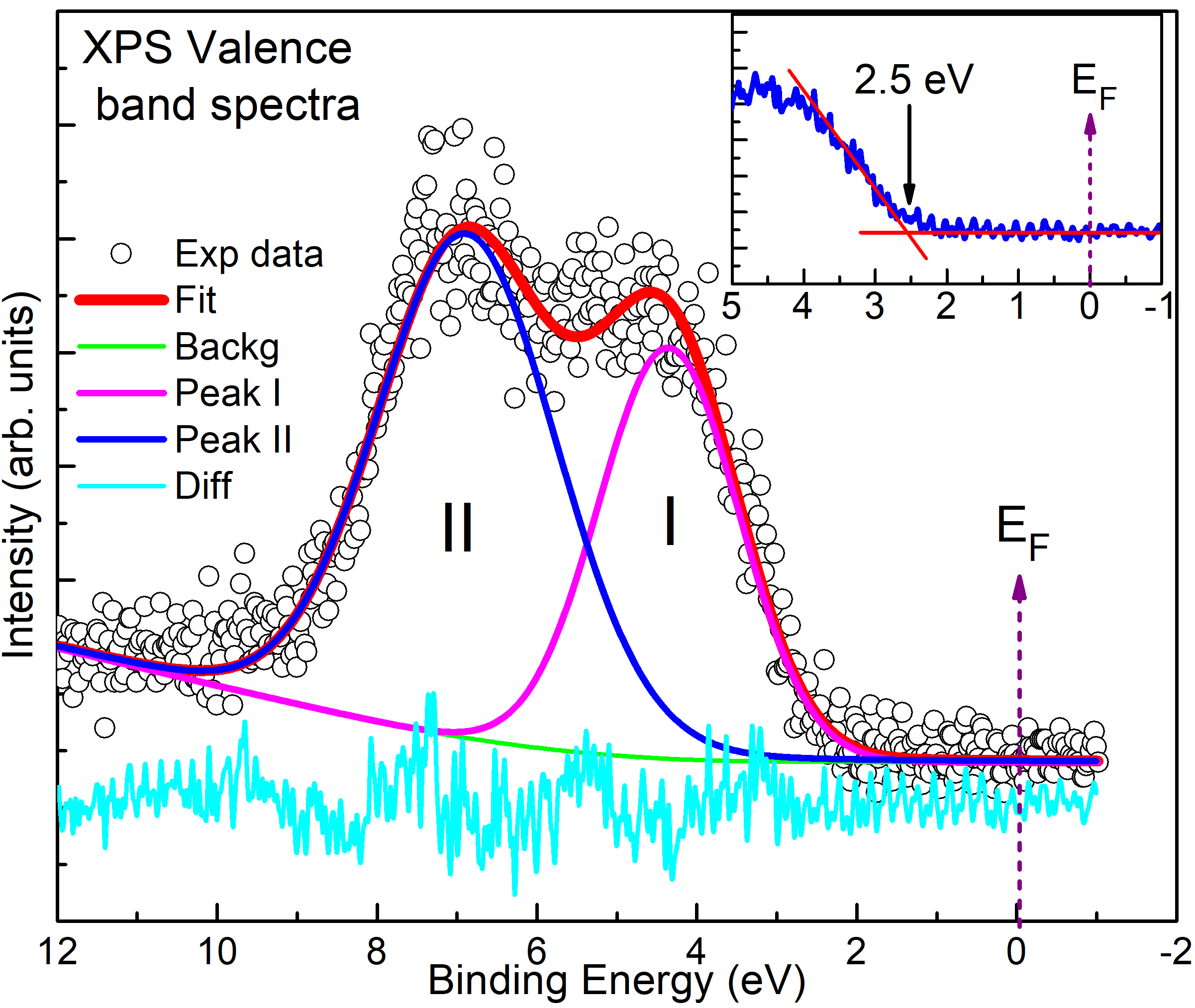}\\
	\caption{Deconvoluted valence-band XPS spectrum of green synthesized TiO$_2$ NPs measured using XPS; top inset shows density of states and bottom inset displays proposed schematic energy band structure derived from VB XPS and UV-vis spectrum.}\label{fig:VBS}
\end{figure}

To determine the occupied and unoccupied density of states (DOS) near Fermi energy level (E$_F$) for green synthesized TiO$_{2-\delta}$ NPs, we measured valence band spectrum (VBS) using XPS shown in Fig.~\ref{fig:VBS}. We have fitted the VBS XPS pectrum using a sum of Gaussian and Lorentzians using two peaks. Peak I (solid pink line) and peak II (solid blue line) are situated at binding energy $\sim$ 4.36~eV and 6.88~eV, respectively. The obtained intensity of Peak I is lower than peak II. The two peaks are associated to some minor contribution from Ti-3d (t$_{2g}$) states (Peak I) and O-2p state (peak II) in the valence band of TiO$_2$. The valence band maximum (VBM) value can be estimated from DOS (following the protocol elsewhere~\cite{Vinod4,Vinod5,Vinod6,Vinod24}) is shown in the top inset of Fig.~\ref{fig:VBS}. It can be estimated by the intersection of fitted linear portion of lower binding energy side and the fitted base line. The estimated value of VBM turns out to be $\sim$ 2.5~eV. Further, the value of conduction band minimum (CBM) can be calculated by the relation CBM = VBM(2.5) - E$_g$(2.75), which is found to be - 0.25~eV. The proposed schematic energy band diagram estimated from the value of E$_g$ and valence band maximum (VBM) is displayed in bottom inset of Fig.~\ref{fig:VBS}.

\begin{figure}
	\centering
	\includegraphics[width=9cm]{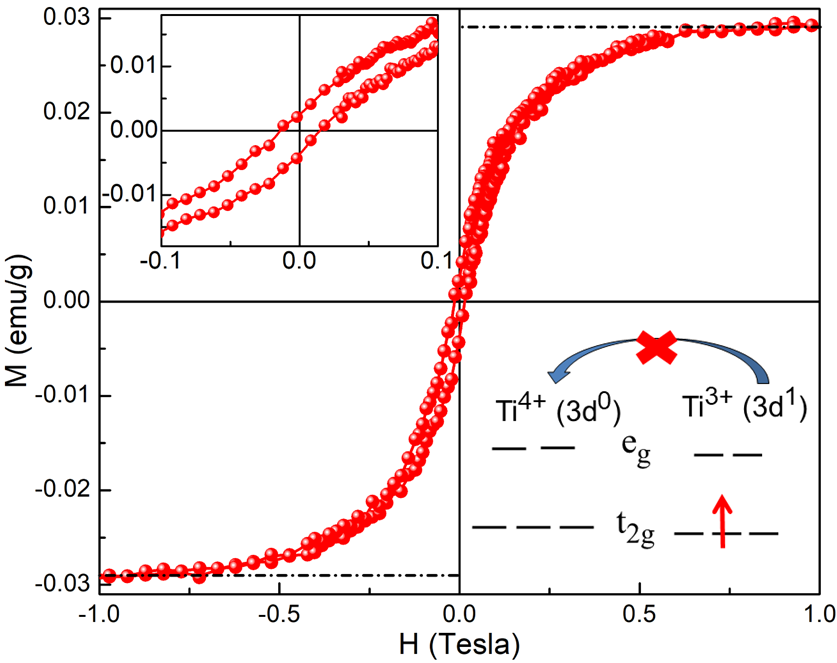}\\
	\caption{Magnetization (M) as a function of applied magnetic field (H) of green synthesized TiO$_{2-\delta}$ NPs; top inset shows enlarged view at low field for the same. Bottom inset reveals a schematic demonstration of spin selective hopping of electrons between Ti${4+}$ and Ti$^{3+}$ combination.}\label{fig:MH}
\end{figure}

Figure~\ref{fig:MH} shows magnetization (M) versus applied magnetic field (H) curve measured at room temperature in a field range -1 to + 1 Tesla for green synthesized TiO$_{2-\delta}$ NPs. Magnetization seems to saturate even at $\pm$ 0.6~T with a clear magnetic hysteresis [top inset Fig.~\ref{fig:MH}], suggests room temperature ferromagnetism. The observed value of saturation magnetization (M$_S$), remanent magnetization (M$_R$) and coercive field (H$_C$) are 0.029~emu/g, 0.003~emu/g and 0.0143~T, respectively. Although, the observed value of M$_S$ for current sample is low, it is relatively {\it{two order higher}} than the reported value in earlier literature for TiO$_2$ NPs synthesized using other plants extract [Table~\ref{Parametercomparison}] and few TiO$_2$ samples prepared by conventional chemical route~\cite{Polo,Patel1,Zhang3}. It is worth to mention that study of magnetic property is {\it{entirely missing}} in previous reports of TiO$_2$ NPs synthesized using PN leaf extract~\cite{Shanavas,Panneerselvam1,Shimi1,Jayan}.

Ideally, TiO$_2$ is nonmagnetic~\cite{Coey2,Hong} due to absence of an unpaired electron in its d-orbital i.e., Ti$^{4+}$(3d$^0$). In current sample, surface oxygen vacancies trap electrons by creating F-center which exhibits two unpaired electrons~\cite{Santara,Zheng}. One of the electrons from F-center tend to occupy the vicinity of Ti$^{4+}$-site by converting it into Ti$^{3+}$(3d$^1$). As a result, electrons can hop from Ti$^{3+}$(3d$^1$)-site to Ti$^{4+}$(3d$^0$)-site [Fig.~\ref{fig:MH} bottom inset]. On the other hand, hopping from Ti$^{4+}$(3d$^0$) to Ti$^{3+}$(3d$^1$)-site is blocked. Therefore, {\it{virtual hopping}} between Ti$^{3+}$(3d$^1$) to Ti$^{4+}$(3d$^0$) state is however, allowed, leading to ferromagnetic double exchange interaction~\cite{Vinod7,Vinod8,Vinod25}. Thus, the larger presence of Ti$^{3+}$-ions due to the higher number of oxygen vacancies is possibly responsible for the enhanced magnetization compared to literatures [Table~\ref{Parametercomparison}].  For current sample, Ti-2p XPS result suggests 13\% contribution from Ti$^{3+}$(3d$^1$)-cation that exhibit M$_S$ $\sim$ 0.00043~$\mu_B$/f.u. (0.029~emu/g), very less than theoretical value 0.225~$\mu_B$ [spin only magnetic moment of Ti$^{3+}$ is 1.73~$\mu_B$]. K. K. Tejas et al.,~\cite{Thejas} found M$_S$ $\sim$ 0.000022~$\mu_B$/f.u. for TiO$_2$ NPs synthesized by chemical route. It is relatively one order less than current sample. Therefore, extremely low value of magnetic moment in TiO$_2$ NPs is not surprising.

\section{CONCLUSIONS}

{\color{red}{}}
We have successfully synthesized oxygen deficient TiO$_{2-\delta}$ nanoparticles (NPs) by green route using {\it{Phyllanthus niruri}} (PN) whole plant extract. Structural characterization using various techniques such as XRD, SEM, EDX, and TEM confirms tetragonal anatase phase formation of TiO$_{2-\delta}$ NPs ($\sim$ 35~nm). In a nutshell, the important observations are : (I) FTIR results confirm the existence of plant derivatives and their functional groups and biomolecules attached to the surface of TiO$_{2-\delta}$ NPs; (II) Ti-2p and O-1s core level XPS confirms presence of surface oxygen vacancies, favors coexistence of mixed oxidation states of Ti i.e., Ti$^{3+}$ and Ti$^{4+}$; (III) a strong absorption peak centered at wavelength $\sim$ 250~nm with relatively lower optical band gap energy E$_g$ = 2.75~eV than bulk polycrystalline TiO$_2$, possibly related to the oxygen vacancies/Ti$^{3+}$ that leads to the formation of donor levels in electronic state of TiO$_{2-\delta}$ NPs; (IV) XPS valence band spectrum confirms the absence of density of states at Fermi energy level E$_F$ with valence band maximum at $\sim$ 2.5~eV; (V) Magnetization result shows room temperature ferromagnetism (M$_S$ $\sim$ 0.029~emu/g, H$_C$ $\sim$ 0.0143~T) with a clear hysteresis. The observed value of M$_S$ is relatively higher (two order magnitude enhancement) compared to TiO$_2$ NPs synthesized using other plants extract. It is worth to mention that magnetic investigation of TiO$_2$ NPs synthesized using PN leaf extract is completely {\it{missing}}. The room temperature ferromagnetism and enhancement of magnetization may be understood by {\it{virtual hopping}} of electrons due to the appearance of surface oxygen vacancies. It decreases the oxidation states of Ti from Ti$^{4+}$ to Ti$^{3+}$ facilitating double exchange mechanism. As a result, electron may hop from Ti$^{3+}$(3d$^1$)-site to Ti$^{4+}$(3d$^0$)-site, however, vice versa is not allowed. We expect that present study [i.e., two order of magnitude enhancement of M$_S$ and significant reduction of E$_g$ compared to TiO$_2$ NPs synthesized by PN leaf extract and other plants] makes PN whole plant extract a better reducing agent for green route synthesis of TiO$_2$ NPs.



\begin{thebibliography}{99}

\bibitem{Rana} A. Rana, S. Pathak, D.-K. Lim, S.-K. Kim, R. Srivastava, S. N. Sharma, R. Verma,
Recent Advancements in Plant-and Microbe-Mediated Synthesis of Metal and Metal Oxide Nanomaterials and Their Emerging Antimicrobial Applications,
\href{https://doi.org/10.1021/acsanm.3c01351}{ACS Appl. Nano Mater. 6 (2023) 2106}.

\bibitem{Chavali} M. S. Chavali, M. P. Nikolova,
Metal oxide nanoparticles and their applications in nanotechnology,
\href{https://doi.org/10.1007/s42452-019-0592-3}{SN Appl. Sci. 1 (2019) 607}.

\bibitem{Negrescu} A. M. Negrescu, M. S. Killian, S. N. V. Raghu, P. Schmuki, A. Mazare, A. Cimpean,
Metal Oxide Nanoparticles: Review of Synthesis, Characterization and Biological Effects,
\href{https://doi.org/10.3390/jfb13040274}{J. Funct. Biomater. 13 (2022) 274}.


\bibitem{Vinod21} A. Juyal, V. K. Dwivedi, S. Verma, S. Nandi, A. Agarwal, and S. Mukhopadhyay,
Possible transition between charge density wave and Weyl semimetal phase in Y$_2$Ir$_2$O$_7$,
\href{https://doi.org/10.1103/PhysRevB.106.155149}{Phys. Rev. B 106 (2022) 155149}.

\bibitem{Vinod22} V. K. Dwivedi, A. Juyal, and S. Mukhopadhyay,
Colossal enhancement of electrical conductivity in Y$_2$Ir$_2$O$_7$ nanoparticles,
\href{https://doi.org/10.1088/2053-1591/3/11/115020}{Mater. Res. Express 3 (2016) 115020}.

\bibitem{Vinod23} V. K. Dwivedi, and S. Mukhopadhyay,
Structural and magnetic studies of nanocrystallinen Y$_2$Ir$_2$O$_7$,
\href{https://doi.org/10.1063/1.4917801}{AIP Conf. Proc. 1665 (2015) 050160}.





\bibitem{Aslam} M. Aslam, A. Z. Abdullah, M. Rafatullah,
Recent development in the green synthesis of titanium dioxide nanoparticles using plant-based biomolecules for environmental and antimicrobial applications,
\href{https://doi.org/10.1016/j.jiec.2021.04.010}{J. Ind. Eng. Chem. 98 (2021) 1}.

\bibitem{Singh} S. Singh, I. C. Maurya, A. Tiwari, P. Srivastava, L. Bahadur,
Green synthesis of TiO$_2$ nanoparticles using {\it{Citrus limon}} juice extract as a bio-capping agent for enhanced performance of dye-sensitized solar cells,
\href{https://doi.org/10.1016/j.surfin.2021.101652}{Surf. Interfaces 28 (2022) 101652}.

\bibitem{Kumar} A. Kumar, A. Chaudhari, S. Kumar, S. Kushwaha,
Biosynthesis of TiO$_2$ nanostructures using {\it{Camellia sinensis}} extract (polyphenols) and investigation of their execution as photoanodes in photovoltaic device,
\href{https://doi.org/10.1016/j.surfin.2024.104154}{Surf. Interfaces 47 (2024) 104154}.

\bibitem{Ullah} A. K. M. A. Ullah, A. N. Tamanna, A. Hossain, M. Akter, M. F. Kabir, A. R. M. Tareq, A. K. M. F. Kibria, M. Kurasaki, M. M. Rahmanc, M. N. I. Khan,
{\it{In vitro}} cytotoxicity and antibiotic application of green route surface modified ferromagnetic TiO$_2$ nanoparticles,
\href{https://doi.org/10.1039/C9RA01395D}{RSC Adv. 9 (2019) 13254}.

\bibitem{Bhullar} S. Bhullar, N. Goyala, S. Gupta,
Rapid green-synthesis of TiO$_2$ nanoparticles for therapeutic applications,
\href{https://doi.org/10.1039/D1RA05588G}{RSC Adv. 11 (2021) 30343}.


\bibitem{Narayanan} M. Narayanan, P. Vigneshwari, D. Natarajan, S. Kandasamy, M. Alsehli, A. Elfasakhany, A. Pugazhendhi,
Synthesis and characterization of TiO$_2$ NPs by aqueous leaf extract of Coleus aromaticus and assess their antibacterial, larvicidal, and anticancer potential,
\href{https://doi.org/10.1016/j.envres.2021.111335}{Environ. Res. 200 (2021) 111335}.

\bibitem{Kashale} A. A. Kashale, K. P. Gattu, K. Ghule, V. H. Ingole, S. Dhanayat, R. Sharma, J.-Y. Chang, A. V. Ghule,
Biomediated green synthesis of TiO$_2$ nanoparticles for lithium ion battery application,
\href{https://doi.org/10.1016/j.compositesb.2016.06.015}{Compos. Part B. Eng. 99 (2016) 297}.

\bibitem{Sargazi} S. Sargazi, S. ER, S. S. Gelen, A. Rahdar, M. Bilal, R. Arshad, N. Ajalli, M. F. A. Khan, S. Pandey,
Application of titanium dioxide nanoparticles in photothermal and photodynamic therapy of cancer: An updated and comprehensive review,
\href{https://doi.org/10.1016/j.jddst.2022.103605}{J. Drug Delivery Sci. Technol. 75 (2022) 103605}.

\bibitem{Lazar} M. A. Lazar, S. Varghese, S. S. Nair,
Photocatalytic Water Treatment by Titanium Dioxide: Recent Updates,
\href{https://doi.org/10.3390/catal2040572}{Catalysts 2 (2012) 572}.

\bibitem{Parras} M. Parras, A. Varela, R. Cortes-Gil, K. Boulahya, A. Hernando, J. M. Gonzalez-Calbet,
Room-Temperature Ferromagnetism in Reduced Rutile TiO$_{2-\delta}$ Nanoparticles,
\href{https://doi.org/10.1021/jz401115q}{J. Phys. Chem. Lett. 4 (2013) 2171}.

\bibitem{Qian} Z. Qian, W. Ping, L. Bao-Ling, L. Zun-Ming J. En-Yong,
Room-Temperature Ferromagnetism in Semiconducting TiO$_{2-\delta}$ Nanoparticles,
\href{https://doi.org/10.1088/0256-307X/25/5/078}{Chinese Phys. Lett. 25 (2008) 1811}.

\bibitem{Dhakshinamoorthy1} J. Dhakshinamoorthy, S. K. Srivastava, D. Mishra, B. Pullithadathil,
Unveiling the interplay between induced native defects and room temperature magnetic ordering in titanium deficient disordered-TiO$_2$ nanoparticles,
\href{https://doi.org/10.1088/1361-6528/abc57b}{Nanotechnology 32 (2021) 095701}.

\bibitem{Dhakshinamoorthy2} J. Dhakshinamoorthy, A. K. Prasad, S. Dhara, B. Pullithadathil,
Anomalous Effects of Lattice Strain and Ti$^{3+}$ Interstitials on Room-Temperature Magnetic Ordering in Defect-Engineered Nano-TiO$_2$,
\href{https://doi.org/10.1021/acs.jpcc.8b09851}{J. Phys. Chem. C 122 (2018) 27782}.



\bibitem{Shi} H. Shi, R. Magaye, V. Castranova, J. Zhao,
Titanium dioxide nanoparticles: a review of current toxicological data,
\href{https://doi.org/10.1186/1743-8977-10-15}{Part. Fibre Toxicol. 10 (2013) 15}.


\bibitem{Mao} X. Chen, S. S. Mao,
Titanium Dioxide Nanomaterials: Synthesis, Properties, Modifications, and Applications,
\href{https://doi.org/10.1021/cr0500535}{Chem. Rev. 107 (2007) 2891}.

\bibitem{Sunny} N. E. Sunny, S. S. Mathew, N. Chandel, P. Saravanan, R. Rajeshkannan, M. Rajasimman, Y. Vasseghian, N. Rajamohan, S. V. Kumar,
Green synthesis of titanium dioxide nanoparticles using plant biomass and their applications- A review,
\href{https://doi.org/10.1016/j.chemosphere.2022.134612}{Chemosphere 300 (2022) 134612}.

\bibitem{Rajaram} P. Rajaram, A. R. Jeice, K. Jayakumar,
Review of green synthesized TiO$_2$ nanoparticles for diverse applications,
\href{https://doi.org/10.1016/j.surfin.2023.102912}{Surf. Interfaces 39 (2023) 102912}.



\bibitem{Azad} A. Azad, H. Zafar, F. Raza, M. Sulaiman,
Factors Influencing the Green Synthesis of Metallic Nanoparticles Using Plant Extracts: A Comprehensive Review,
\href{https://doi.org/10.1055/s-0043-1774289}{Pharmaceutical Fronts 5 (2023) e117}.




\bibitem{Lee1} N. Y. S. Lee, W. K. S. Khoo, M. A. Adnan, T. P. Mahalingam, A. R. Fernandez, K. Jeevaratnam,
The pharmacological potential of {\it{Phyllanthus niruri}},
\href{https://doi.org/10.1111/jphp.12565}{J. Pharm. Pharmacol. 68 (2016) 953}.

\bibitem{Bagalkotkar} G. Bagalkotkar, S. R. Sagineedu, M. S. Saad, J. Stanslas,
Phytochemicals from Phyllanthus niruri Linn. and their pharmacological properties: a review,
\href{https://doi.org/10.1211/jpp.58.12.0001}{J. Pharm. Pharmacol. 58 (2006) 1559}.

\bibitem{Khanna} A. K. Khanna, F. Rizvi, R. Chander,
Lipid lowering activity of {\it{Phyllanthus niruri}} in hyperlipemic rats,
\href{https://doi.org/10.1016/S0378-8741(02)00136-8}{J. Ethnopharmacol. 82 (2002) 19}.

\bibitem{Mishra} L. Mishra, V. K. Dwivedi, H. K. Dara, V. K. Chakradhary, S. Ithineni, A. G. Prabhudessai, S. Nehar,
Core/Shell-Like Magnetic Structure and Optical Properties in CuO Nanoparticles Synthesized by Green Route,
\href{https://doi.org/10.1021/acssusresmgt.4c00325}{ACS Sustainable Resour. Manage. 1 (2024) 2472}.



\bibitem{Shanavas} S. Shanavas, A. Priyadharsan, S. Karthikeyan, K. Dharmaboopathi, I. Ragavan, C. Vidya, R. Acevedo, P. M. Anbarasana,
Green synthesis of titanium dioxide nanoparticles using {\it{Phyllanthus niruri}} leaf extract and study on its structural, optical and morphological properties,
\href{https://doi.org/10.1016/j.matpr.2019.06.715}{Mater. Today: Proc. 26 (2020) 3531}.

\bibitem{Panneerselvam1} A. Panneerselvam, J. Velayutham, S. Ramasamy,
Green synthesis of TiO$_2$ nanoparticles prepared from {\it{Phyllanthus niruri}} leaf extract for dye adsorption and their isotherm and kinetic studies,
\href{https://doi.org/10.1049/nbt2.12033}{IET Nanobiotechnol. 15 (2021) 164}.

\bibitem{Shimi1} A. K. Shimi, S. M. Wabaidur, M. R. Siddiqui, M. A. Islam, K. P. Rane, T. S. A. Jeevan,
Photocatalytic Activity of Green Construction TiO$_2$ Nanoparticles from {\it{Phyllanthus niruri}} Leaf Extract,
\href{https://doi.org/10.1155/2022/7011539}{J. Nanomater. 2022 (2022)}.

\bibitem{Jayan} N. Jayan, L. D. B. Metta,
Process optimization by response surface methodology-central composite design for the adsorption of lead by green synthesized TiO$_2$ using {\it{Phyllanthus acidus}} extract,
\href{https://doi.org/10.1007/s13399-022-02534-w}{Biomass Conv. Bioref. 14 (2024) 825}.



\bibitem{Sarkar} A. Sarkar, G. G. Khan,
The formation and detection techniques of oxygen vacancies in titanium oxide-based nanostructures,
\href{https://doi.org/10.1039/C8NR09666J}{Nanoscale 11 (2019) 3414}.


\bibitem{Swamy} V. Swamy, A. Kuznetsov, L. S. Dubrovinsky, R. A. Caruso, D. G. Shchukin, B. C. Muddle,
Finite-size and pressure effects on the Raman spectrum of nanocrystalline anatase TiO$_2$,
\href{https://doi.org/10.1103/PhysRevB.71.184302}{Phys. Rev. B 71 (2005) 184302}.

\bibitem{Parker2} J. C. Parker, R. W. Siegel,
Calibration of the Raman spectrum to the oxygen stoichiometry of nanophase TiO$_2$,
\href{https://doi.org/10.1063/1.104274}{Appl. Phys. Lett. 57 (1990) 943}.

\bibitem{Kumari} A. Kumari, A. Kumar, R. Dawn, J. Roy, S. Jena, R. Vinjamuri, D. Panda, S. K. Sahoo, V. K. Verma, S. Mahapatra, A. Rahaman, A. Ahlawat, M. Gupta, K. Kumar, A. Kandasami, V. R. Singh,
Effect of annealing temperature on the structural, electronic and magnetic properties of Co doped TiO$_2$ nanoparticles: An investigation by synchrotron-based experimental techniques,
\href{https://doi.org/10.1016/j.jallcom.2022.167739}{J. Alloys Compd. 933 (2023) 167739}.


\bibitem{Sanjines} R. Sanjines, H. Tang, H. Berger, F. Gozzo, G. Margaritondo, F. Levy,
Electronic structure of anatase TiO$_2$ oxide,
\href{https://doi.org/10.1063/1.356190}{J. Appl. Phys. 75 (1994) 2945}.

\bibitem{Thejas} K. K. Thejas, K. K. Supin, V. R. Akshay, B. Arun, G. Mandal, A. Chanda, M. Vasundhara,
Effect of annealing time on structural, optical and magnetic properties of TiO$_2$ nanoparticles,
\href{https://doi.org/10.1016/j.optmat.2022.113178}{Opt. Mater. 134 (2022) 113178}.

\bibitem{Kumari1} A. Kumari, W. W. Tjiu, Z. Aabdin, J. Roy, V. K. Verma, A. Kandasami, V. R. Singh,
Valence band spectroscopic study of Co-doped TiO$_2$ nanoparticles using synchrotron based advanced spectroscopic techniques,
\href{https://doi.org/10.1016/j.apsusc.2023.157732}{Appl. Surf. Sci. 635 (2023) 157732}.





\bibitem{Li2} H. Li, S. Wang, J. Tang, H. Xie, J. Ma, H. Chi, C. Li,
Roles of oxygen vacancies in surface plasmon resonance photoelectrocatalytic water oxidation,
\href{https://doi.org/10.1016/j.xcrp.2023.101386}{Cell Rep. Phys. Sci. 4 (2023) 101386}.

\bibitem{Li3} S. Li, X. Bian, J. Gao, G. Zhu, M. Hojamberdiev, C. Wang, X. Wei,
Effect of oxygen vacancy and surface plasmon resonance: a photocatalytic activity study on Ag/Bi$_4$Ti$_3$O$_{12}$ nanocomposites,
\href{https://doi.org/10.1007/s10854-017-7731-7}{J. Mater. Sci.: Mater. Electron. 28 (2017) 17896}.

\bibitem{Yang1} B. Yang, Z. Ma, Q. Li, X. Liu, Z. Liu, W. Yang, X. Guo, X. Jia,
Regulation of surface plasmon resonance and oxygen vacancy defects in chlorine doped Bi-BiO$_{2-x}$ for imidacloprid photocatalytic degradation,
\href{https://doi.org/10.1039/C9NJ04936C}{New J. Chem. 44 (2020) 1090}.

\bibitem{Yang2} L. Yang, X. Jiang, W. Ruan, B. Zhao, W. Xu, J. R. Lombardi,
Observation of Enhanced Raman Scattering for Molecules Adsorbed on TiO$_2$ Nanoparticles: Charge-Transfer Contribution,
\href{https://doi.org/10.1021/jp8074145}{J. Phys. Chem. C 112 (2008) 20095}.

\bibitem{Pan} X. Pan, M.-Q. Yang, X. Fu, N. Zhang, Y.-J. Xu,
Defective TiO$_2$ with oxygen vacancies: synthesis, properties and photocatalytic applications,
\href{https://doi.org/10.1039/C3NR00476G}{Nanoscale 5 (2013) 3601}.

\bibitem{Song} H. Song, C. Li, Z. Lou, Z. Ye, L. Zhu,
Effective Formation of Oxygen Vacancies in Black TiO$_2$ Nanostructures with Efficient Solar-Driven Water Splitting,
\href{https://doi.org/10.1021/acssuschemeng.7b01774}{ACS Sustainable Chem. Eng. 5 (2017) 8982}.

\bibitem{Yu} X. Yu, Y. Li, X. Hu, D. Zhang, Y. Tao, Z. Liu, Y. He, Md. A. Haque, Z. Liu, T. Wu, Q. J. Wang,
Narrow bandgap oxide nanoparticles coupled with graphene for high performance mid-infrared photodetection,
\href{https://doi.org/10.1038/s41467-018-06776-z}{Nat. Commun. 9 (2018) 4299}.

\bibitem{Vijeta} V. Singh, J. J. Pulikkotil,
Electronic phase transition and transport properties of Ti$_2$O$_3$,
\href{https://doi.org/10.1016/j.jallcom.2015.10.203}{J. Alloys Compd. 658 (2016) 430}.





\bibitem{Scanlon} D. O. Scanlon, C. W. Dunnill, J. Buckeridge, S. A. Shevlin, A. J. Logsdail, S. M. Woodley, C. R. A. Catlow, M. J. Powell, R. G. Palgrave, I. P. Parkin, G. W. Watson, T. W. Keal, P. Sherwood, A. Walsh, A. A. Sokol,
Band alignment of rutile and anatase TiO$_2$,
\href{https://doi.org/10.1038/nmat3697}{Nature Mater. 12 (2013) 798}.

\bibitem{Dey} B. Dey, S. K. Panda, J. Mallick, S. Sen, B. N. Parida, A. Mondal, M. Kar, S. K. Srivastava,
Observation of room temperature d$^0$ ferromagnetism, bandgap narrowing, zero dielectric loss, dielectric enhancement in highly transparent p-type Na-doped rutile TiO$_2$ compounds for spintronics applications,
\href{https://doi.org/10.1016/j.jallcom.2022.167442}{J. Alloys Compd. 930 (2023) 167442}.


\bibitem{Dette} C. Dette, M. A. P.-Osorio, C. S. Kley, P. Punke, C. E. Patrick, P. Jacobson, F. Giustino, S. J. Jung, K. Kern,
TiO$_2$ Anatase with a Bandgap in the Visible Region,
\href{https://doi.org/10.1021/nl503131s}{Nano Lett. 14 (2014) 6533}.


\bibitem{Vinod4} V. K. Dwivedi, S. Mukhopadhyay,
Influence of electronic structure parameters on the electrical transport and magnetic properties of Y$_{2-x}$Bi$_x$Ir$_2$O$_7$ pyrochlore iridates,
\href{https://doi.org/10.1063/1.5125254}{J. Appl. Phys. 126 (2019) 165112}.

\bibitem{Vinod5} V. K. Dwivedi, S. Mukhopadhyay,
Suppression of long range magnetic ordering and electrical conduction in Y$_{1.7}$Bi$_{0.3}$Ir$_2$O$_7$ thin film,
\href{https://doi.org/10.1016/j.jmmm.2019.04.049}{J. Magn. Magn. Mater. 484 (2019) 313}.

\bibitem{Vinod6} V. K. Dwivedi, S. Mukhopadhyay,
Study of optical properties of polycrystalline Y$_2$Ir$_2$O$_7$,
\href{https://doi.org/10.1063/1.4980569}{AIP Conf. Proc. 1832 (2017) 090016}.

\bibitem{Vinod24} V. K. Dwivedi , and S. Mukhopadhyay,
Evolution of structural, magnetic and electrical transport properties of Y$_{1.7}$Bi$_{0.3}$Ir$_2$O$_7$ thin film grown on YSZ(100) substrate by PLD,
\href{https://doi.org/10.1016/j.physb.2019.07.006}{Phys. B: Condens. Matter 571 (2019) 137}.



\bibitem{Polo} C. G.-Polo, S. Larumbe, J. M. Pastor,
Room temperature ferromagnetism in non-magnetic doped TiO$_2$ nanoparticles,
\href{https://doi.org/10.1063/1.4795615}{J. Appl. Phys. 113 (2013) 17B511}.

\bibitem{Patel1} S. K. S. Patel, S. Kurian, N. S. Gajbhiye,
Room-temperature ferromagnetism of Fe-doped TiO$_2$ nanoparticles driven by oxygen vacancy,
\href{https://doi.org/10.1016/j.materresbull.2012.11.031}{Mater. Res. Bull. 48 (2013) 655}.

\bibitem{Zhang3} H. Zhang, W. Huang, R. Lin, Y. Wang, B. Long, Q. Hu, Y. Wu,
Room temperature ferromagnetism in pristine TiO$_2$ nanoparticles triggered by singly ionized surface oxygen vacancy induced via calcining in different air pressure,
\href{https://doi.org/10.1016/j.jallcom.2020.157913}{J. Alloys Compd. 860 (2021) 157913}.


\bibitem{Coey2} J. M. D. Coey, 
Magnetism in d$^0$ oxides,
\href{https://doi.org/10.1038/s41563-019-0365-9}{Nat. Mater. 18 (2019) 652}.

\bibitem{Hong} N. H. Hong, J. Sakai, N. Poirot, V. Brize, 
Room-temperature ferromagnetism observed in undoped semiconducting and insulating oxide thin films,
\href{https://doi.org/10.1103/PhysRevB.73.132404}{Phys. Rev. B 73 (2006) 132404}.

\bibitem{Santara} B. Santara, P. K. Giri, K. Imakitab, M. Fujii,
Evidence of oxygen vacancy induced room temperature ferromagnetism in solvothermally synthesized undoped TiO$_2$ nanoribbons,
\href{https://doi.org/10.1039/C3NR00799E}{Nanoscale 5 (2013) 5476}.

\bibitem{Zheng} J.-Y. Zheng, S.-H. Bao, Y.-H. Lv, P. Jin,
Activation and Enhancement of Room-Temperature Ferromagnetism in Cu-Doped Anatase TiO$_2$ Films by Bound Magnetic Polaron and Oxygen Defects,
\href{https://doi.org/10.1021/am506013w}{ACS Appl. Mater. Interfaces 6 (2014) 22243}.



\bibitem{Vinod7} V. K. Dwivedi, S. Mukhopadhyay,
Coexistence of high electrical conductivity and weak ferromagnetism in Cr doped Y$_2$Ir$_2$O$_7$ pyrochlore iridates,
\href{https://doi.org/10.1063/1.5100316}{J. Appl. Phys. 125 (2019) 223901}.

\bibitem{Vinod8} V. K. Dwivedi, 
Identification of a Griffiths-like phase and its evolution in the substituted pyrochlore iridates Y$_2$Ir$_{2-x}$Cr$_x$O$_7$ (x = 0.0, 0.05, 0.1, 0.2),
\href{https://doi.org/10.1103/PhysRevB.107.134413}{Phys. Rev. B 107 (2023) 134413}.

\bibitem{Vinod25} V. K. Dwivedi, P. Mandal, and S. Mukhopadhyay,
Frustration-Induced Inversion of the Magnetocaloric Effect and Metamagnetic Transition in Substituted Pyrochlore Iridates,
\href{https://doi.org/10.1021/acsaelm.1c01294}{ACS Appl. Electron. Mater. 4 (2022) 1611}.




\bibitem{Hong34} N. H. Hong, J. Sakai, N. Poirot, and V. Brize, 
Room-temperature ferromagnetism observed in undoped semiconducting and insulating oxide thin films,
\href{https://doi.org/10.1103/PhysRevB.73.132404}{{Phys. Rev. B} {\bf{73}}, 132404 (2006)}.








						
\end{thebibliography}
\end{document}